\documentstyle[12pt]{article}
\def\lapproxeq{\lower .7ex\hbox{$\;\stackrel{\textstyle <}{\sim}\;$}}
\def\gapproxeq{\lower .7ex\hbox{$\;\stackrel{\textstyle >}{\sim}\;$}}

\begin{document}

\titlepage

\begin{flushright} DTP/93/64 \\ August 1993
\end{flushright}

\begin{center}
\vspace*{2cm}
{\large{\bf How suppressed are the radiative interference effects \\ in heavy
unstable particle production? }}
\end{center}

\vspace*{.75cm}
\begin{center}
V.S.\ Fadin\footnote{Permanent address: Budker Institute for Nuclear Physics
and Novosibirsk State University, 630090 Novosibirsk, Russia.}, V.A.\ Khoze
and A.D.\ Martin \\
Department of Physics, University of Durham, \\ Durham DH1 3LE, England.
\end{center}

\vspace*{2cm}

\begin{abstract}
We present a \lq\lq theorem" which quantifies, for inclusive processes, the
level of suppression of the radiative interference effects between the
production and decay stages of heavy {\it unstable} particles.  The theorem,
which is based on very general physical arguments, is applicable to all orders
in the coupling.
\end{abstract}

\newpage

The study of heavy unstable particles (e.g.\ $t$ quark or $W$ boson) is of
particular interest in high energy physics.  Since the typical width of such a
particle is large, $\Gamma \sim O$(1 GeV), the particle is not observed
itself, but rather it is identified by its decay products.  Therefore in
production processes of heavy unstable particles it is natural to separate the
production stage from the decay processes.  In general these stages are not
independent and may be interconnected by radiative interference effects.
Particle(s) (e.g.\ photon(s) and/or gluon(s)) could be produced at one stage
and absorbed at another; we speak of virtual interference.  Real interference
will occur as well since the same real particle can be emitted from the
different stages of the process.

Many observations rely on a clear understanding of the role of these
interference effects.  Indeed there is a long list of examples where a
detailed knowledge of the effects can be important for the interpretation of
experimental data (see, for example, refs.\ \cite{FKM,A1}). For instance, the
interference phenomena have to be quantified to obtain the complete
$O(\alpha_s)$ corrections to $e^+e^- \rightarrow t\bar{t} \rightarrow
W^+W^-b\bar{b}$ and the $O(\alpha^2_s)$ corrections to $e^+e^- \rightarrow
W^+W^- \rightarrow$ (4 jets) inclusive cross sections.  (Note that the leading
QCD interference correction to the latter process is $O(\alpha^2_s)$ to ensure
colour conservation.)  Clearly such corrections are relevant, interalia, for
an accurate experimental determination of heavy particle masses.

In ref.\ \cite{FKM} we made an explicit study, with the help of soft-insertion
techniques,  of the interference phenomena between the production and decay
stages to $O(\alpha_{{\rm int}})$ (where $\alpha_{{\rm int}} = \alpha$ or
$\alpha_s$ as appropriate) and, in each case, we identified the particular
degree of inclusiveness of the process that is required for the interference
effects to be suppressed.  Of course, the real and virtual interference
contributions, taken separately, are not suppressed but are infrared
divergent.  Clearly the infrared divergent parts have to cancel for physically
meaning values.  This is not the issue.  Rather the crucial question is to
what level does this cancellation occur?

Here we present a general theorem which states that the effects of radiative
interferences (between the various production and decay stages) are each
suppressed by $O(\Gamma/M)$ in the totally inclusive production of unstable
particles.  Our proof does not rely on specific assumptions, like
soft-insertion factorization, and the resulting theorem is applicable to any
order in the coupling $\alpha_{{\rm int}}$.

Before presenting the proof it may be useful to express the resulting recipe
in a symbolic form.  To be specific let us consider the production of $N$
heavy unstable particles $A_1,...A_N$ with masses $M_i$ and widths $\Gamma_i$.
For reference purposes we first consider the production in the absence of the
decays of the heavy particles (i.e.\ we switch off the interaction responsible
for the decays).  Then the inclusive cross section may be written in the form
\begin{equation}
\sigma_{{\rm stable}}(A_1,...A_N) \;\; = \;\; \sigma_0(M^2_i) (1 +
\delta(R,C)) ,
\end{equation}
where $\sigma_0(M^2_i)$ is the production cross section in the Born
approximation and $\delta(R,C)$ represents the radiative corrections.  Here we
have separated possible Coulombic corrections $C$ from the remaining radiative
corrections $R$.  The Coulombic effects $C$, which are associated with large
space-time intervals, are only important \cite{A3} if two charged (or
coloured) particles are slowly moving\footnote{Of course the separation of
Coulomb effects can only be done uniquely near threshold, but this is the very
region where the instability effects are most important.} in their c.m.\ frame
(e.g.\ $W^+W^-$ or $t\bar{t}$ production near threshold).  Now the question is
\lq\lq how is (1) modified in the physical case when we allow the heavy
unstable particles $A_i$ to decay?"  We will show
\begin{equation}
\sigma_{{\rm unstable}}(A_1,...A_N) \;\; = \;\; \int \prod_i \left(
ds_i\rho(s_i) \right) \sigma_0(s_i) (1 + \delta(R,\bar{C})) + \sum_n O \left(
\alpha^n_{{\rm int}} \frac{\Gamma_i}{M_i} \right)
\end{equation}
with
\begin{equation}
\rho(s_i) \;\; = \;\; \frac{\sqrt{s_i} \, \Gamma_i(s_i)}{\pi[(s_i-M^2_i)^2 +
s_i\Gamma^2_i (s_i)]}
\end{equation}
where $\Gamma_i(s_i)$ is \lq\lq running" physical width which incorporates the
radiative effects associated solely with the decay of particle $A_i$.

In other words, the theorem for the production of heavy {\it unstable}
particles says that, apart from the two modifications explicitly shown in the
formula, the introduction of the particle widths gives rise to no new
corrections up to order $\alpha_{{\rm int}}^n\Gamma_i/M_i$, where the $n = 0$
term corresponds to the standard (non-radiative) non-resonant
backgrounds\footnote{Strictly speaking the notation $\sigma(A_1,...A_N)$ in
formula (2) is imprecise in the sense that the same final state may be reached
without going through resonances $A_1,...A_N$.} and the $n \geq 1$ terms to
interference induced by $n$ radiated quanta.  The two modifications in going
from (1) to (2) are, first, the natural kinematic effect leading to the
integrations over $\rho(s_i)$ and, second, the modification (symbolically
shown as $C \rightarrow \bar{C}$) of the Coulombic interaction between
particle pairs which are non-relativistic in their c.m.\ frame \cite{A2,FKM1}.
In particular, the theorem says the remaining radiative corrections are
unchanged.

The proof of the theorem relies on two facts.  First, the suppression (by at
least a factor $\Gamma/|k^0|$) of interferences between the various production
and decay stages arising from energetic photons/gluons with $|k^0| \gg
\Gamma$, and, second, the absence of infrared divergences in the total or
inclusive cross-section.  Both facts have a simple physical interpretation.
Let us consider, without loss of generality, the case when the total energy of
the process is of the order of the masses $M_i \sim M$ of the produced
unstable particles.  Then the typical time, $\tau_p$, of the duration of the
production stage, as well as of the decay stages, is of order of $1/M$.  This
time $\tau_p$ is much less than the characteristic time $\Delta t$ between the
various stages; since the typical time between the production and decay stages
of any heavy particle $i$ is $\Delta t = \Delta t_i \sim 1/\Gamma_i \sim
1/\Gamma$, and that between the decay stages of heavy particles $i$ and $j$ is
\begin{displaymath}
\Delta t \;\; = \;\; \Delta t_{ij} \sim {\rm  max} \left( \frac{1}{\Gamma_i},
\frac{1}{\Gamma_j} \right) \sim \frac{1}{\Gamma}.
\end{displaymath}
As a consequence the relative phases of photon/gluon emissions (or between
emission and absorption) at the different stages of the process are
approximately equal to $|k^0| \Delta t \sim |k^0|/\Gamma$.  When $|k^0| \gg
\Gamma$ this phase shift is large and therefore the radiative interference
effects are suppressed.

This physical picture is readily confirmed by inspection of the propagators of
the unstable particles.  Photon/gluon emission with energy $|k^0| \gg \Gamma$
shifts the invariant mass of the unstable particle far from its resonant
value, so for an interference contribution  we obtain a factor $\sim
1/(i\Gamma M)(k^0M - i\Gamma M)$, instead of the factor $\sim |1/i\Gamma M|^2$
in the absence of radiation.  Thus we have suppression of hard photon/gluon
interference effects by at least a factor of $\Gamma/|k^0|$.

The second basic fact, the absence of infrared divergences in totally
inclusive cross-sections \cite{BN}, has also a clear physical interpretation.
Infrared divergences appear when we use states containing a definite number of
photons/gluons.  Now the acceleration of charge/colour leads to radiation with
finite spectral intensity at zero frequency and so the scattering or the
creation of charged/coloured particles is accompanied by the emission of an
infinite number of photons/gluons.  To obtain a physically meaningful cross
section (which can be expanded in powers of $\alpha$ or $\alpha_s$) we must
include arbitrary numbers of emitted photons or gluons.  In the case of gluons
we need also to average over the initial colour states, because colour is
changed in the process of gluon emission.

For the proof we only need the absence of those infrared divergences in the
total cross-section which are connected with the radiative interference
between the various production and decay stages of the process.  To show this
we first note that the total cross-section is proportional to the imaginary
part of the forward scattering amplitude.  When this amplitude is expressed as
the sum of Feynman diagrams, all particles except the initial particles,
appear as internal lines, that is they correspond to virtual particles.
Therefore the interference photon/gluon lines must be attached to internal
lines, at least at one end.  But it is well known (and can be readily verified
by simple power counting) that infrared divergent contributions only arise
from photons/gluons with lines which couple at both ends to external lines
corresponding to on-mass-shell particles.  Because of the absence of infrared
divergences the contribution from the region of small photon/gluon energies
$|k^0| \lapproxeq \Gamma$ is small due to the lack of phase space.  On the
other hand, as shown above, for $|k^0| \gg \Gamma$ the interference effects
are small due to the large time separations between the various stages.
Therefore radiative interference between the production and decay stages is
suppressed by at least a factor $\Gamma/M$.

So far we have considered interference induced by a single photon or gluon.
For multiple exchanges the situation is slightly different but the conclusion
remains valid.  In the case of multiple exchange large energies $k^0_i$ of
individual photons/gluons are allowed, since the constraint that the invariant
mass of the unstable particle must not be shifted far from its resonant value
only requires
\begin{equation}
\left| \sum_i  k^0_i \right| \lapproxeq \Gamma
\end{equation}
where the sum is performed over photons or gluons emitted ($k^0_i > 0$) and
absorbed ($k^0_i < 0$) at one of the decay stages.  But because of the absence
of infrared divergences there is again a suppression of at least one factor of
$\Gamma/M$ due to the restriction of the phase space imposed by condition (4).

Finally we must consider the Coulomb radiative effects.  Now there are
infrared singularities connected with the Coulomb interaction of
charged/coloured particles which are special in the sense that they are not
cancelled by real emissions (which have only transverse polarization states),
but rather they appear in matrix elements as a phase factor with an infinite
phase.  Therefore they do not appear in the expression for the cross section
and are not seen at all in our approach where we express the cross section in
terms of the forward scattering amplitude.

However there is a Coulomb interaction which is connected with small
photon/gluon frequencies, but which is not infrared divergent.  For two
charged/coloured particles, with reduced mass $\mu$ and momentum
$\mbox{\boldmath $q$}$ in their c.m.\ frame, the essential energies $k^0_c$ of
the exchanged Coulombic photons/gluons are typically $|k^0_c| \sim
\mbox{\boldmath $q$}^2/\sqrt{(\mbox{\boldmath $q$}^2 + \mu^2)}. \,\,$  When
$\mbox{\boldmath $q$}^2 \lapproxeq \mu\Gamma$ these energies are $|k^0_c|
\lapproxeq \Gamma$.  Thus for two slowly moving charged/coloured particles in
their c.m.\ frame there is an important Coulombic radiative interaction coming
from the region of small photon/gluon energies.  At first sight this appears
to violate our previous statement, that the contribution from the region
$|k^0| \lapproxeq \Gamma$ is small.   But that statement referred to
interference photons/gluons and it remains correct for them.  The reason is
that for interference between the different production and decay stages of a
process, the only important Coulomb interactions are those between a
charged/coloured decay product of one of the unstable particles and some other
particle (e.g.\ another unstable particle or one of its decay products) with
the interacting pair slowly moving in their c.m.\ frame (that is
$\mbox{\boldmath $q$}^2 \lapproxeq \mu\Gamma$).  This corresponds to a very
small region of the available phase space and so these Coulomb effects are
also suppressed by at least a factor $\Gamma/M$.  This concludes the proof of
the theorem.

Some words should be added to explain the modification $\delta(R,C)
\rightarrow \delta(R,\bar{C})$ in going from formula (1) for \lq\lq stable"
heavy particles to the realistic formula (2) for unstable particles.  Away
from the heavy particle production threshold the typical heavy particle
interaction time is $1/\sqrt{s} \lapproxeq 1/M$, i.e.\ much smaller than their
lifetimes.  Thus the influence of instability on the Coulomb corrections at
the production stage gives effects of relative order $\Gamma/M$ or less.  Thus
$\delta(R,\bar{C}) \approx \delta(R,C)$.

The situation is different for heavy unstable particle production near
threshold.  Then the typical Coulomb interaction time $\tau_c$ can be
comparable to, or even larger than, the particle's lifetime $\tau$
\begin{equation}
\tau_c \; \sim \; \frac{1}{k^0_c} \; \sim \; \frac{\mu}{\mbox{\boldmath
$q$}^2} \; \gapproxeq \; \frac{1}{\Gamma} \;\; = \;\; \tau .
\end{equation}
Therefore the Coulomb part $C$ of the radiative correction $\delta$ shown in
(1) will be considerably modified by instability, and hence it is denoted by
$\bar{C}$ in (2).  The calculation \cite{A2,FKM1} of the modified contribution
$\bar{C}$ can be best done using old-fashioned non-relativistic perturbation
theory, particularly as these effects are only crucial near threshold.  The
diagrams are the same as in the stable particle case, but we require the
momentum $p_i = (\epsilon_i,\mbox{\boldmath $p$}_i)$ of $A_i$ to satisfy
$p^2_i = s_i$, and we must replace the energies of the unstable particles by
\begin{equation}
\epsilon \; \longrightarrow \; \epsilon +
\frac{iM\Gamma}{2\sqrt{\mbox{\boldmath $p$}^2 + M^2}}
\end{equation}
in the energy denominators of all intermediate states.

In conclusion, formula (2) gives  a transparent recipe for quantifying the
level of suppression of the radiative interference effects (which occur
between the production and decay stages) in the inclusive production of heavy
{\it unstable} particles.  Moreover we detail how the modifications due to
instability may be implemented (i.e.\ the introduction of $\rho(s_i)$
containing the {\it physical} width $\Gamma(s_i)$ and the modifications $C
\rightarrow \bar{C}$ of the Coulomb effects near threshold).

\vspace*{1cm}
\noindent {\bf Acknowledgements}
\vspace*{.5cm}

This work was supported in part by the U.K.\ Science and Engineering Research
Council.  V.S.F.\ thanks the Centre for Particle Theory at the University of
Durham for hospitality.  V.A.K.\ is grateful to Yu.L.\ Dokshitzer for useful
discussions.

\end{document}